\documentclass[letterpaper, 10 pt, conference]{ieeeconf}  

\IEEEoverridecommandlockouts                              
\usepackage{graphicx}
\usepackage{lipsum}  
\usepackage{xcolor}
\usepackage{fancyhdr}

\usepackage{url}
\usepackage{dirtree}
\usepackage{listings}
\usepackage{cite}
\usepackage{hyperref}
\usepackage{cleveref}

\lstdefinestyle{jsonieee}{
    basicstyle=\ttfamily\footnotesize, 
    numbers=none,
    breaklines=false,
    frame=none,
    backgroundcolor=\color{white},
    showstringspaces=false,
    tabsize=2,
    captionpos=b,
    xleftmargin=0pt,
    xrightmargin=0pt,
    aboveskip=5pt,
    belowskip=2pt
}

\graphicspath{ {images/} }

\Crefname{figure}{Fig.}{Figures}

\title{A Cross-Perspective Annotated Dataset for Dynamic Object-Level Interest Modeling in Cloud Gaming }

\author{
  Hongqin Lei \\
  Nanjing University of Posts \\
  and Telecommunications, Nanjing\\
  \texttt{lhg8945@gmail.com} \\
  \and
  Haowei Tang \\
  Nanjing University of Posts \\
  and Telecommunications, Nanjing\\
  \texttt{b20010530@njupt.edu.cn} \\
  \and
  Zhe Zhang \\
  Nanjing University of Posts\\
  and Telecommunications, Nanjing\\
  \texttt{zhezhang@njupt.edu.cn} \\
}

\begin{document}

\maketitle  
\thispagestyle{plain}

\begin{abstract}
Cloud gaming has gained popularity as it provides high-quality gaming experiences on thin hardware, such as phones and tablets. 
Transmitting gameplay frames at high resolutions and ultra-low latency is the key to guaranteeing players' quality of experience (QoE).
Numerous studies have explored deep learning (DL) techniques to address this challenge. 
The efficiency of these DL-based approaches is highly affected by the dataset.
However, existing datasets usually focus on the positions of objects while ignoring semantic relationships with other objects and their unique features.
In this paper, we present a game dataset by collecting gameplay clips from Grand Theft Auto (GTA) V, and annotating the player's interested objects during the gameplay.
Based on the collected data, we analyze several factors that have an impact on player's interest and identify that the player's in-game speed, object's size, and object's speed are the main factors.
The dataset is available at \href{https://drive.google.com/drive/folders/1idH251a2K-hGGd3pKjX-3Gx5o_rUqLC4?usp=sharing}{https://drive.google.com/drive/folders/1idH251a2K-hGGd3pKjX-3Gx5o\_rUqLC4?usp=sharing}

\end{abstract}

\section{Introduction}\label{sec::intro}

Traditional high-quality games require high-performance local devices, which limits the accessibility of ordinary players. Cloud gaming could reduce the demand for local graphics processing units (GPUs) and enable high-quality games on low-specification devices, thus attracting interest from players. The advancement of real-time communication technologies has further promoted this trend. As a result, it has led high-tech companies to launch their cloud gaming services, such as Nvidia's GeForce NOW \cite{nvidia}, Sony's PlayStation service \cite{playstation}, and Microsoft's Xbox service \cite{xbox}. Cloud gaming service providers leverage the powerful GPUs on cloud servers to render game content and transmit gameplay scenes to players \cite{huang2014gaminganywhere}.




Cloud gaming has stringent requirements in terms of bandwidth and latency. To address this, many methods have been proposed, such as adaptive bitrate streaming, scheduling policy, and video coding. 
Main video encoders just compress videos by minimizing temporal and spatial redundancy based on image changes. Recent studies consider the subjectivity of visual perception. They employ deep learning (DL) methods to extract regions of interest (ROI) and then compress the video. These demonstrate priority for interested objects in the scene. For instance, Xue \emph{et al}. extracts ROIs from video conference through DL methods and delivers different quantization parameters (QPs) to ROIs and Non-ROIs to enhance portrait quality \cite{portaitROI}. 
The accuracy of DL-based methods depends on high-quality datasets. Most of the gaming datasets define objects as key objects if: 1). they are at the center of the scene; 2). they occupy more than half of the scene. Such a definition ignores the unique features of objects and the semantic relationships with other objects. They annotate ROI by bounding boxes instead of object-level annotations. Besides, When playing action-oriented games like action role-playing games (ARPGs) and open-world action-adventure games (OWAAGs), it is clear that the player's in-game speed plays a critical role in the distribution of interested objects. Previous works ignore players' in-game speed. Consequently, the extraction of ROIs from these datasets proves to be relatively straightforward.


Motivated by the above challenge, existing datasets do not support object-level ROI, encoding based on unique features, semantic relationships, and the player's in-game speed.  We create a cross-perspective gaming dataset with dynamic object-level annotations. In GTA V, gameplay scenes are similar to the real world. The player's interest and behavior constantly change due to variations of cross-perceptions, which include the player's in-game speed, the unique features of objects, and the semantic relationships with other objects.

The novel dataset in this paper is a collection of typical scenarios from GTA V, designed to support the extraction of fine-grained interested objects. The dataset comprises 501 video clips and 1503 game images from GTA V. Each image corresponds to 2 annotation JSON files. We then respectively analyze the factors that influence cross-perception. This further distinguishes the main factors and the secondary factors.

Compared to existing cloud gaming datasets, the dataset in this paper has the following significant features:

\noindent\textbf{Varying Scenes:} Three scenarios are categorized based on player's in-game speed: stationary, low speed, and high speed. Varying speed has a significant influence on classes of interested objects, which have been neglected in previous gaming datasets.

\noindent\textbf{Multi-Interest:} The annotations for each image are generated by combining the interests from 5 different observers. Each image in the dataset contains one or more interested objects with annotations. Compared to single-interest, multi-interest annotations are more likely to reflect the diversity among players and ensure stable video coding based on interests.

\noindent\textbf{Cross-Perception:} Through analysis of the collected dataset,  we classify the factors influencing interest into main and secondary factors. The main factors include the player's in-game speed, the object's size, and the object's speed. Secondary factors are color contrast and the object's shape. Experiments on the player's speed indicate that the distribution of the object's class will vary significantly at different speeds.

\section{Related Work}\label{sec::Related Work}

\subsection{Cloud Gaming Datasets}\label{subsec::Cloud Gaming Video Datasets}

Recent studies have developed numerous gaming datasets, covering various types of games.
Barman \emph{et al}. in \cite{HDR-gaming-video-streaming} discussed the performance of various coding tools on gaming content with high dynamic range (HDR) and ultra high definition (UHD) resolutions, which are becoming more prevalent with the rise of cloud gaming services.
Datasets presented in \cite{zhao2021videodataset} collected multiplayer online battle games (e.g., Arena of Valor and Fortnite). It highlighted that objects in first-person games have rich affine motion characteristics. Authors also applied the dataset to existing video coding tools and evaluated their performance.
A large-scale game affect dataset was constructed in \cite{melhart2022arousal}, aiming to investigate the generality of affective computing and directly map pixels to motion by DL methods.
In \cite{GameVideoSET}, raw videos from twelve popular games were collected. The author used H.264 to encode the raw game video at 15 resolution-bitrate pairs, and analyzed the encoding results of different pairs by subjective and objective quality assessment metrics.
To produce audio that matches the game graphics for developers with limited budgets, a novel game audio dataset was proposed in \cite{NES-Video-Musi}. It collected videos of 389 games from the Nintendo Entertainment System and separated the audio from the videos. Game developers utilize neural generative models to rapidly generate audio prototypes based on game videos, thereby guiding the final soundtrack.
In \cite{Stereoscopic-3D-Dataset} Kirill \emph{et al}. developed a mod that can synthesize stereoscopic or multi-angle video datasets with geometric distortion from GTA V. These distortions can cause discomfort when watching 3D videos. This paper trained a convolutional neural network on this dataset to detect distortion in stereoscopic videos.


\begin{figure*}[t!]
    \centering
    \includegraphics[width=1\linewidth]{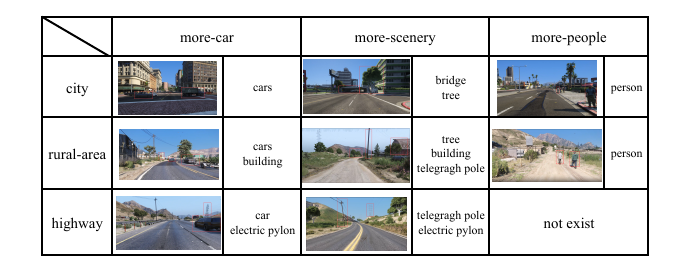}
    \caption{Example images of different scenes from the dataset, and the corresponding multi-interest annotation for each image are marked at the right of the related image.}
    \label{fig:showImages}
    \vspace{-10pt}
\end{figure*}

\begin{figure*}[t!]
    \centering
    \includegraphics[width=1\linewidth]{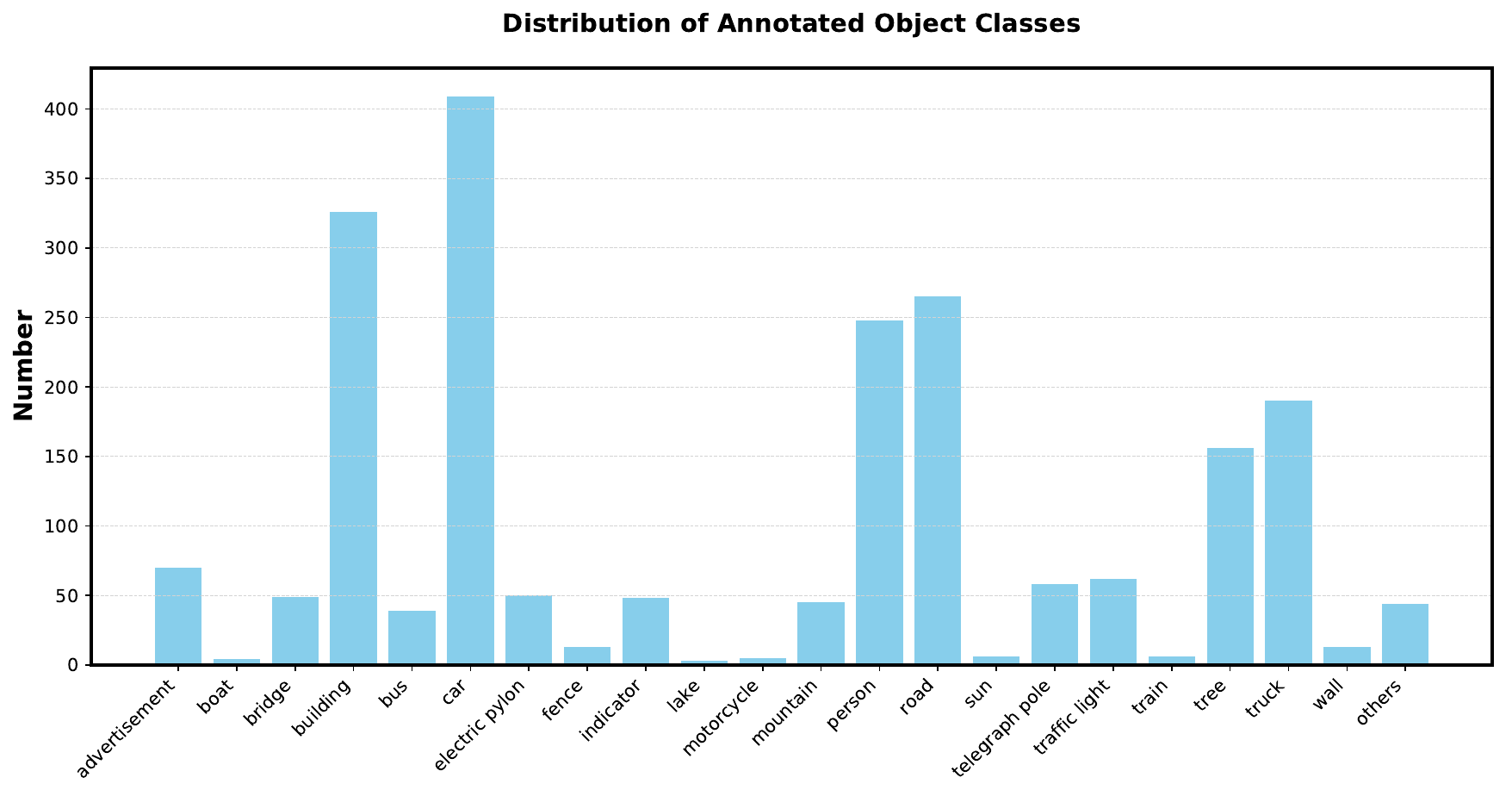}
    \caption{Illustration of the sample numbers of 22 predefined classes in the dataset.}
    \label{fig:labelcounts}
    \vspace{-10pt}
\end{figure*}

\subsection{Effective Video Encoders}\label{subsec:: Effective Video Coding}

H.264 has been widely adopted due to its extensive hardware and software support \cite{codec:h264+overview1, codec:h264+wireless}.
As the successor to H.264, H.265 has more diverse intra-frame and inter-frame prediction modes \cite{codec:hevcbook, codec:hevc+inter, codec:hevc+intra}. These enhanced prediction techniques reduce spatial and temporal redundancy, allowing the encoder to improve video quality at the same bitrate.
There are also several video encoders developed specifically for video streaming. 
VP9 was developed by Google as an alternative to HEVC with considerable efficiency \cite{codec:vp9}. VP9 has low coding complexity and hardware decoding support, which makes it stable on web video and mobile devices \cite{codec:vp9+web-mobile}. It has been a popular choice for platforms like YouTube and Chrome.
AV1 was developed by alliance for open media (AOMedia) and is adopted by major streaming platforms such as Netflix and YouTube. The compression efficiency of AV1 is 23\%-30\% higher than VP9, and the encoding time overhead is 55-58 times higher than VP9 \cite{codec:Future-video-coding-technology, codec:Comparison-of-compression}. Given the significant performance gains of AV1, the trade-off is considered acceptable. Specifically, AV1 has the best performance compared to previous encoders on UHD-HDR content \cite{HDR-gaming-video-streaming}.
As the latest generation of video coding standards, advanced audio coding (AAC), also known as H.266, can improve the compression efficiency of about 50\% than HEVC, and greatly reduce the file size under the same picture quality, which is suitable for 4K/8K UHD video transmission \cite{HDR-gaming-video-streaming}.

As DL technology makes advances in computer vision \cite{cheng2021maskformer}, a variety of studies are exploring how to utilize visual models to predict ROI in video. By encoding ROIs and non-ROIs with different video parameters, it is possible to reduce the bandwidth required for video transmission while maintaining visual quality.
Existing ROI prediction methods can be roughly divided into two categories. One category leverages object detection and classification \cite{ROI-coding-Frank-Wolfe, ROI-DVC, ROI:RL+gaming+coding, ROI:XMligong, ROI:silency-guided-ROI}, to identify ROIs by calculating the degree of interest.
The other approaches directly predict pixel-level ROIs through video saliency prediction \cite{Silency:High-definition-video,silency:real-time2025,silency:springer2023}. These approaches highly rely on eye-tracking datasets and complicated computer vision models.


\section{Data Description and Collection}\label{sec::Data Description and Collection}
\subsection{Data Description}\label{subsec::Data Description}
The dataset in this paper comprises 501 3-minute video clips and 1503 images from GTA V. \Cref{fig:showImages} depicts several typical scenarios in the game. Each image corresponds to two annotation JSON files. One contains manually annotated information about the interested objects as shown in \Cref{fig:json-interest}, and the other contains all objects in an image as shown in \Cref{fig:json-all}.

The dataset is categorized into three levels based on player's in-game speed: stationary (denoted by speed0), low speed (denoted by speed1), and high speed (denoted by speed2). Within each speed level, the scenario is further categorized into city, rural area, and highway. For each speed level and scenario diversity, the dominant visual elements are categorized into high pedestrian density (denoted by more-people), high vehicle density (denoted by more-car), and rich scenery features (denoted by more-scenery), during video collection. It is worth noticing that high pedestrian density and highway scenario type naturally clash. Each scene is reviewed by five independent observers and is annotated with their interest in the gameplay scenes. One or more interested objects within an image are annotated, which is referred to as \textit{multi-interest}. \Cref{fig:showImages} displays the diversity of scenes and the corresponding multi-interest annotations. A total of 22 object classes are annotated for the interested objects. The distribution of object classes is depicted in \Cref{fig:labelcounts}. Significant variations exist among class frequencies. \textit{car, building, people, road, tree}, and \textit{trucks} are the most commonly annotated interested objects as they are the most frequent and relevant in the context of driving.

\begin{figure}[ht]
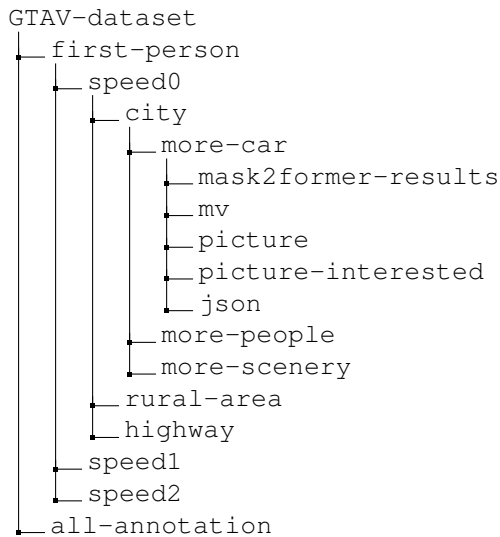

\centering
\begin{minipage}{1\linewidth}
\begin{minipage}{\linewidth}
\dirtree{%
  .1 GTAV-dataset.
  .2 first-person.
  .3 speed0.
  .4 city.
  .5 more-car.
  .6 mask2former-results.
  .6 mv.
  .6 picture.
  .6 picture-interested.
  .6 json.
  .5 more-people.
  .5 more-scenery.
  .4 rural-area.
  .4 highway.
  .3 speed1.
  .3 speed2.
  .2 all-annotation.
}
\end{minipage}
\end{minipage}
\caption{Directory structure of the GTAV-dataset.}
\label{fig:gtav-dir}
\end{figure}

The structure of the dataset is outlined in \Cref{fig:gtav-dir}. In each specific category, there are five subdirectories. Among these subdirectories, the ``video'' folder contains 3-second videos collected from GTA V. The ``picture'' folder includes images extracted from these videos. Interested objects are manually annotated with rectangular bounding boxes, and the annotated images are stored in the ``picture-interest'' folder. The visualization results of semantic segmentation are stored in the ``mask2former-results'' folder. Additionally, each annotated image corresponds to two JSON files: one JSON file containing information about the manually annotated interested objects is placed in the ``int-annotation'' directory, and another JSON file containing information about all objects in the image is stored in the second-level ``all-annotation'' folder.

\begin{figure}[t!]
\centering
\begin{minipage}{1\textwidth}
\begin{lstlisting}[style=jsonieee]
{
  "file_name": [{
      "left_top": {"x": <int>, "y": <int>},
      "right_bottom": {"x": <int>, "y": <int>},
      "label": <string>,
      "size": <int>,
      "distance": <int>,
      "score": <int>,
      "speed": <int>
  }, 
  {object2}]
}
\end{lstlisting}
\end{minipage}
\caption{JSON fields about manually annotated interested objects.}
\label{fig:json-interest}
\end{figure}

\begin{figure}[htbp]
\centering
\begin{minipage}{1\textwidth}
\begin{lstlisting}[style=jsonieee]
[
    {
      "segment_id": <int>,
      "category": <string>,
      "interest": <int: 0 or 1>
      "position": {
         "center": {"x": <int>, "y": <int>},
         "left_top": {"x": <int>, "y": <int>},
         "right_bottom": {"x": <int>, "y": <int>}
       },
       "size": <int>,
       "center_distance": <int>,
       "player_distance": <int>,
       "motion_vector": 
       {"x": <float>, "y": <float>},
       "play_speed": <float: 0 or 0.5 or 1>
    },
    {object2}
]
\end{lstlisting}
\end{minipage}
\caption{JSON fields about overall objects in an image.}
\label{fig:json-all}
\vspace{-10pt}
\end{figure}
 
The detailed information about the interested objects manually annotated by bounding boxes is recorded in the JSON files in \Cref{fig:json-interest}. The fields in the JSON files describe these objects, where \texttt{label} represents the object class, \texttt{left\_top}, \texttt{right\_bottom}, and \texttt{size} indicate the position and size of the bounding box. \texttt{distance} is calculated by the distance from the center of the bounding box to the player's position. The midpoint of the lower edge of the image is viewed as the player's position in this paper.
The \texttt{score} and \texttt{speed} fields are subjective annotations, where a higher \texttt{score} indicates a relatively stronger focus priority, and a larger \texttt{speed} corresponds to a relatively faster speed. 
Furthermore, the detailed information about overall objects is based on semantic segmentation results using mask regions as shown in \Cref{fig:json-all}. The information is object-level, providing specific details for each object in the image. \texttt{category} represents the object class. Compared to the JSON file about interested objects, the JSON structure about overall objects removes the subjectively annotated \texttt{score} and \texttt{speed}. Instead, we introduce two additional perceptual information: \texttt{center\_distance} and \texttt{motion\_vector}, representing the distance from the image center and the object's speed, respectively. \texttt{player\_distance} is calculated by the distance from the center of the object's ``mask'' from semantic segmentation to the player's position. It also includes a \texttt{segment\_id} field, which is automatically assigned by a script and has no special meaning.

\subsection{Data Collection}\label{subsec::data collection}
The dataset is created by collecting driving scenes from GTA V. Following the category criteria described in \Cref{subsec::Data Description}, we collect 1-minute video segments for each category, resulting in a total of 24 1-minute video segments.

To facilitate the annotation of interested objects from the videos, we perform data processing on the collected 1-minute gameplay segments.
First, the 1-minute segment is divided into 20 smaller 3-second video clips. Then we manually extract 3 frames with visual differences from each 3-second video clip for further processing.
Second, five observers watch each three-second video clip and annotate interested objects by bounding boxes on the corresponding frames. Most of the bounding box annotations contain the entire object, indicating that the object is of interest. However, objects that occupy a large portion of the image are not suitable for annotation in this way. In such cases, the bounding box only annotates a part of the object as ROI.
Third, we employ the Mask2Former \cite{cheng2021maskformer} model on the pre-trained COCO dataset \cite{coco2014microsoft} to perform object-level annotation. This model segments each object in the frames and generates a corresponding segmented mask for each. Then we annotate each object with interest value by combining the segmented mask generated with the manually annotated bounding box positions. This integration ensures that the interest value is assigned accurately.
Fourth, we use the dense optical flow \cite{farneback2003two} method to estimate the \texttt{motion\_vector} within the mask regions. By applying this method to the segmented mask regions, we obtain the object's speed.
Finally, to analyze what factors of an object are of interest to people, the attention score of each is manually marked from 5 to 1, which reflects the focus priorities of different objects in an image. 

\section{Analysis}\label{sec::Analysis}

Upon analyzing and reviewing videos in datasets repeatedly, we have identified three main factors and two secondary factors that have a significant impact on players' interest. The main factors are the player's in-game speed, the object's size, and the object's speed. The secondary factors are the object's color and the rarity of the object.

\subsection{Main Factors}\label{subsec:: main factors}
\subsubsection{Impacts of the player's in-game speed on classes of Interested Objects}\label{subsubsec::player's in-game speed}



In immersive games, players' speed considerably influences interested objects. For example, in first-person driving games, the interested objects dramatically shift depending on the player's in-game speed. Higher speed requires a higher level of proficiency and reaction time \cite{information_attention1_2011search, information_att2_2013attention}. The player will pay more attention to the information to avoid a crash and maintain control of the vehicle.

\begin{figure}[t!]
    \centering
    \includegraphics[width=0.75\linewidth]{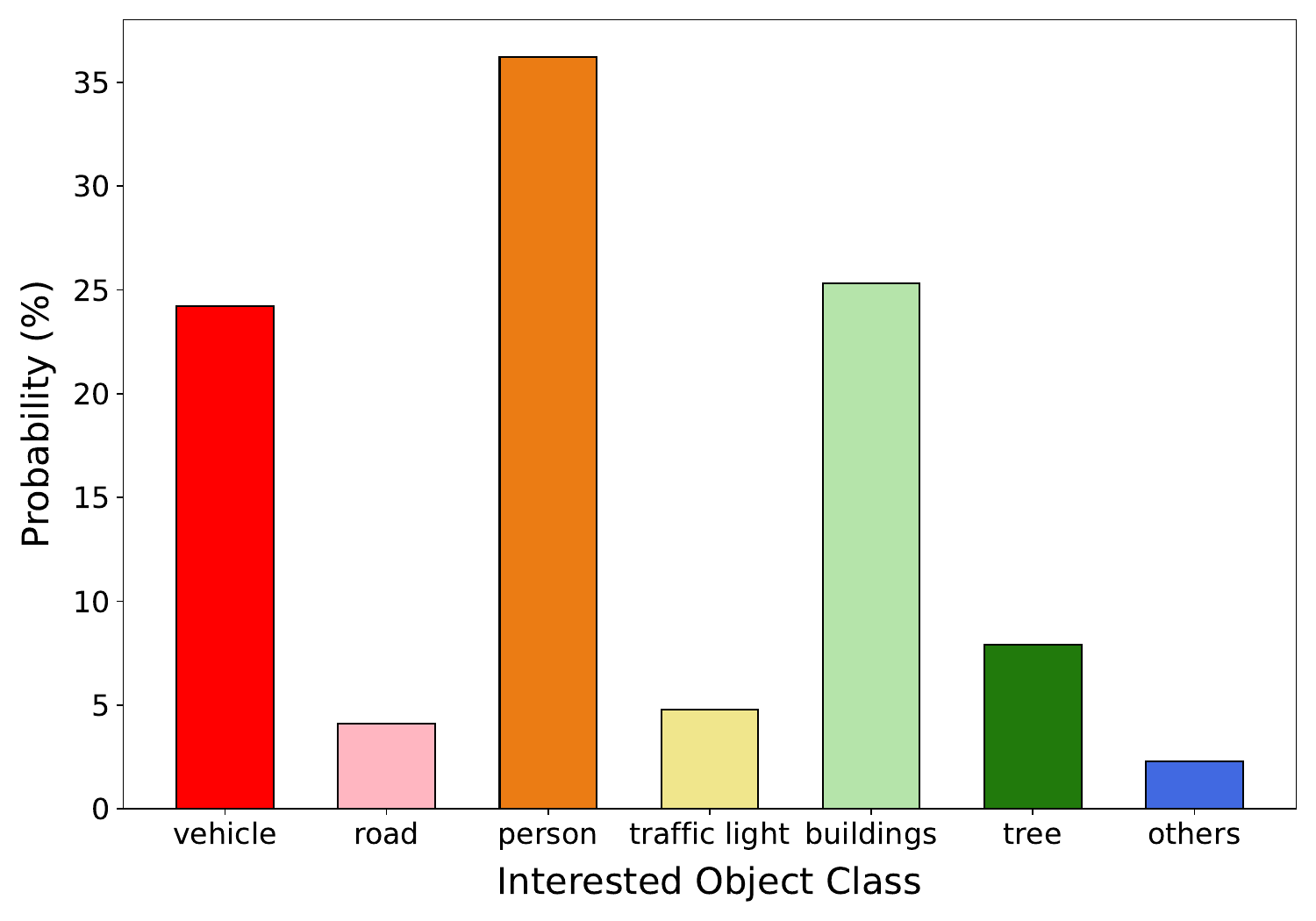}
    \caption{Probability distribution of the player's interested objects in stationary states.}
    \label{fig:stationary}
\end{figure}

\begin{figure}
    \centering
    \includegraphics[width=0.75\linewidth]{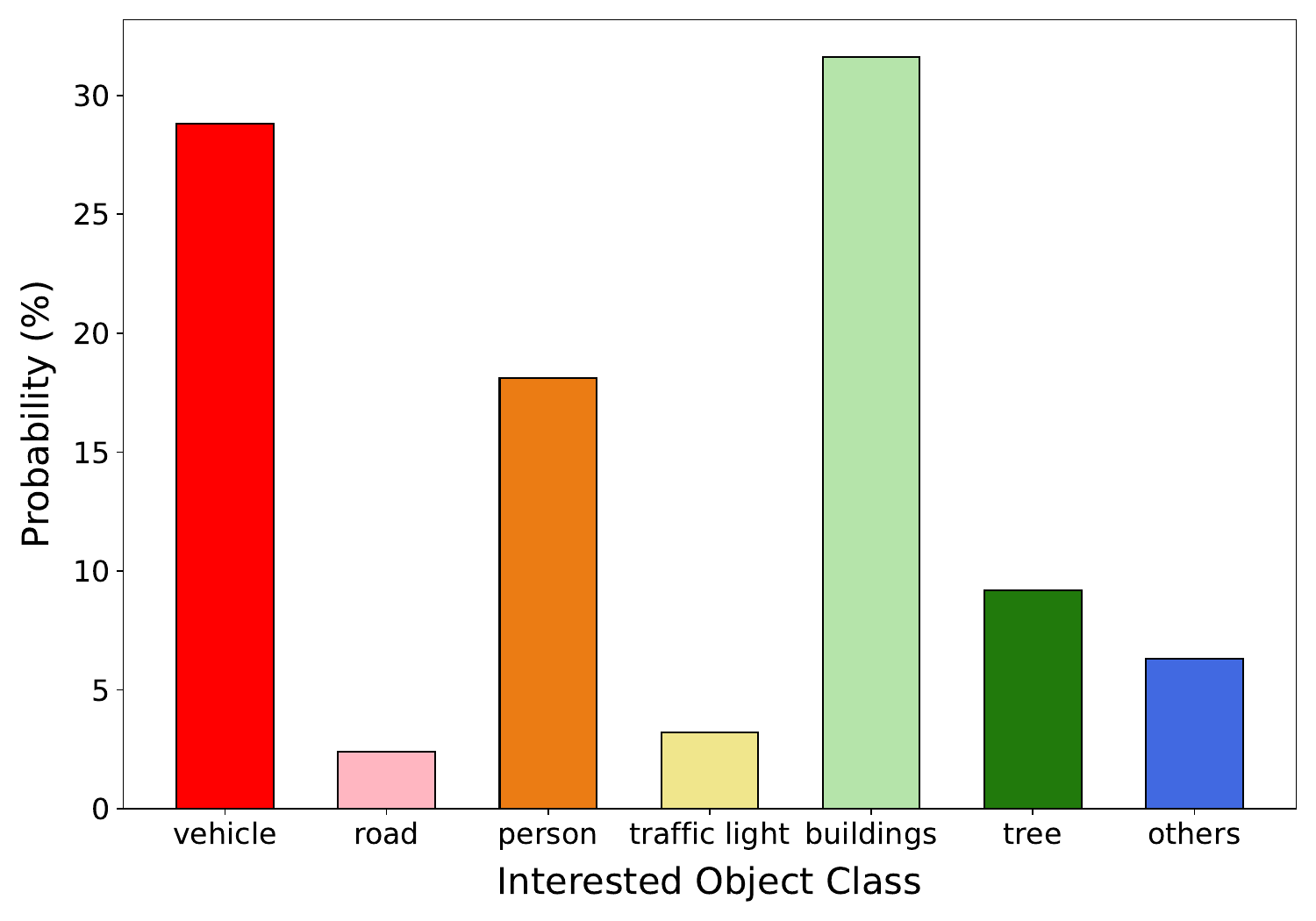}
    \caption{Probability distribution of the player's interested objects in low-speed states.}
    \label{fig:low speed}
\end{figure}

\begin{figure}[ht]
    \centering
    \includegraphics[width=0.75\linewidth]{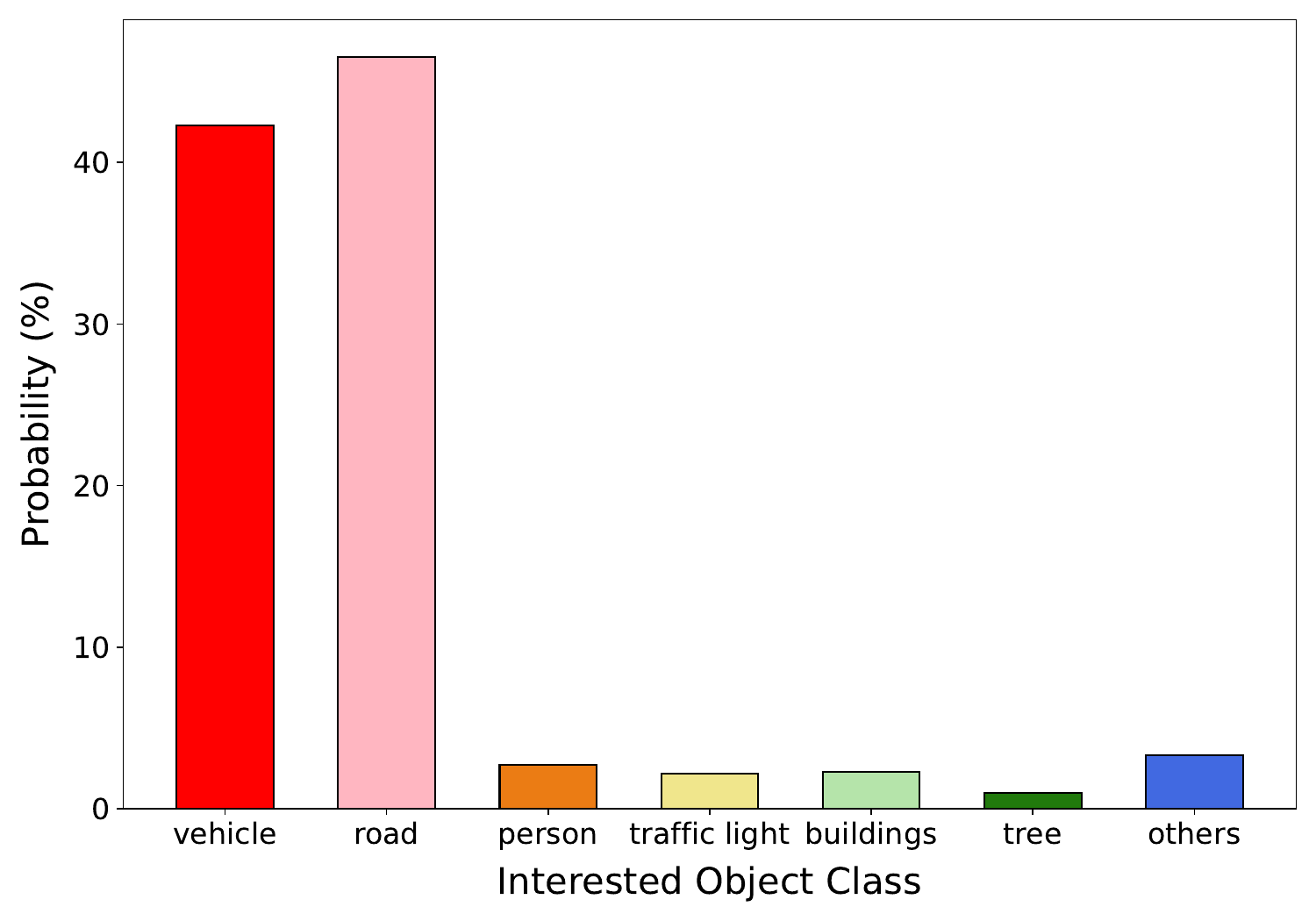}
    \caption{Probability distribution of the player's interested objects in high-speed states.}
    \label{fig:high speed}
\end{figure}

\begin{figure}
    \centering
    \includegraphics[width=0.9\linewidth]{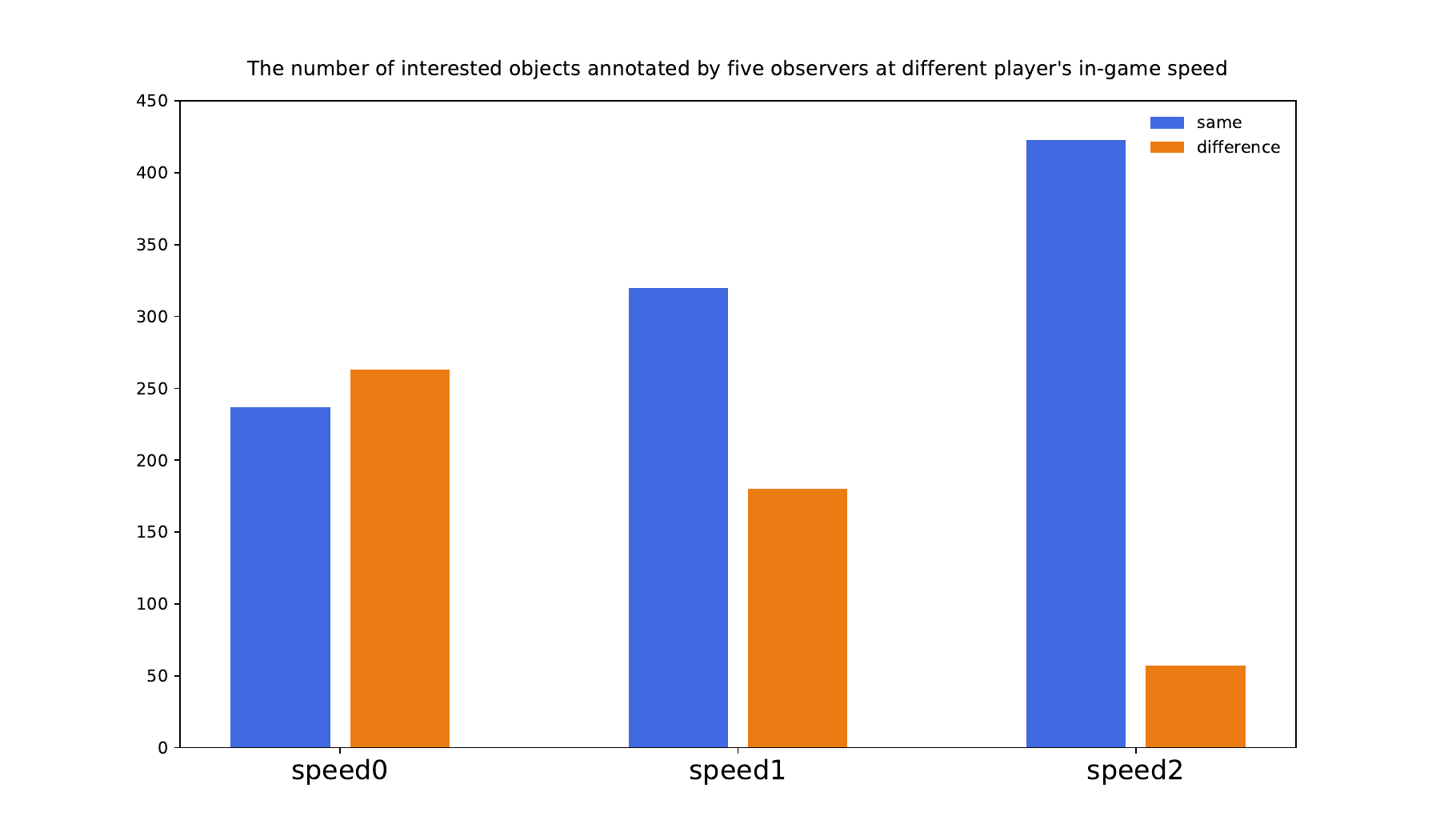}
    \caption{The number of consistent and inconsistent interested objects among five observers at different player's in-game speeds.}
    \label{fig:samediff}
\end{figure}

\begin{figure*}[t!]
    \centering
    \includegraphics[width=1\linewidth]{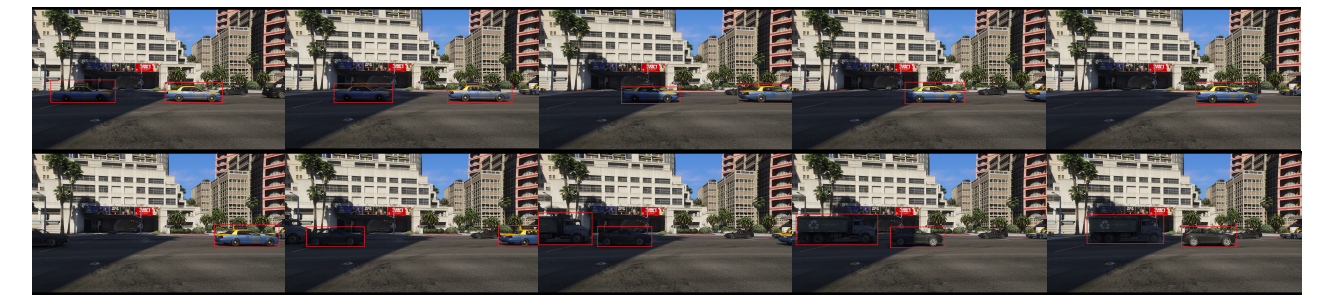}
    \caption{An example of dynamic objects attracting focus.}
    \label{fig:dynamic}
\end{figure*}

\begin{figure}[ht]
    \centering
    \includegraphics[width=0.8\linewidth]{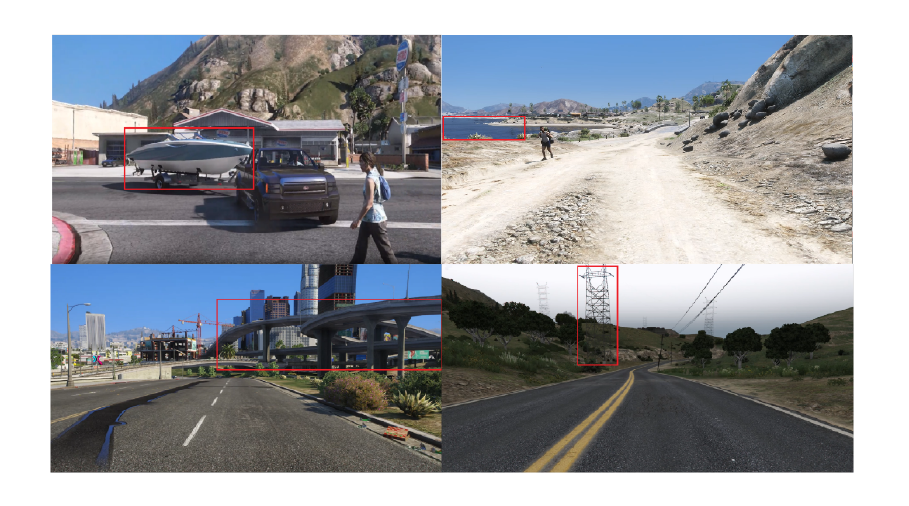}
    \caption{The top row of images shows that objects appear in scenes where they do not typically belong. The bottom row shows objects with uncommon shapes.}
    \label{fig:shape}
\end{figure}

The probability distribution of the player’s interested objects in stationary states is shown in \Cref{fig:stationary}. The distribution of interested objects is similar to the low-speed state, which is illustrated in \Cref{fig:low speed}. Different classes of objects vary greatly in attracting players' interest. Notably, vehicles, people, and buildings account for a larger proportion of interested objects compared to other classes. \Cref{fig:high speed} illustrates the distribution of player’s interested objects in high-speed states. In contrast to stationary and low-speed states, players in high-speed states tend to pay more attention to the road surface and vehicles on the road because it’s necessary to prevent crashes and keep the vehicle steady.

The phenomenon where increasing player's in-game speed reduces the diversity of interested object classes is called interest aggregation. To illustrate the phenomenon more directly, we introduce a simple metric: the number of consistent and inconsistent interested objects across five observers. Five new observers watch video games with varying player's in-game speed and label the interested objects. As shown in \Cref{fig:samediff}, observers tend to watch different objects in stationary and low-speed states. However, as speed increases, observers tend to focus on specific objects, which are always the road surface and vehicles on the road.

\subsubsection{Impacts of Object's Size and Distance}\label{subsubsec::size and distance}

In immersive games, the object's size and distance are important factors in attracting players' attention. 

The distance from the player to objects influences whether the player is focused on them. For objects that originally possess the ability to attract the player, the closer the distance from the object to the player, the greater the probability of attraction.

The pixel-based distance calculation in 2D images ignores the perspective relationships in 3D space, leading to inaccurate distance measurements. Objects closer to the player in 3D space may appear farther away in the 2D image, especially when the horizontal pixel distance exceeds the vertical distance, causing distorted distance values.
To address this problem, we introduce the object's size that can effectively represent the distance from the player to objects in a 2D image. 
We obtain the object's size by counting how many pixels it has. The size accurately reflects the distance from the player and the object in a 2D image. For instance, when a small car is very close to the player, its larger size better reflects its distance to the player.





\subsubsection{Impacts of Object’s Speed}\label{subsubsec::motion states}

The object's speed is an important factor in attracting players' interest. 
When the player is in a stationary or low-speed state, the object's speed becomes a key factor influencing the player's focus. Objects often loom and disappear from the screen quickly, which tends to draw the player's focus.
Conversely, when the player is at high speed, the impact of the object's speed on focus is reduced. Maintaining the stability of the vehicle becomes a priority. Players are inclined to focus on the road ahead to avoid crashing with sudden appearances of new vehicles.
The conclusion from \cite{Moving-looming-stimuli-capture-attention} demonstrates that moving vehicles are more likely to draw the player's interest compared to other objects (e.g., buildings, trees, and stationary vehicles). Particularly moving vehicles that have just entered the gameplay screen are attractive than those disappearing from the screen. We also get a similar conclusion, where the abrupt appearance of a new object and its motion attract individuals. As shown in \Cref{fig:dynamic}, vehicles moving away from the game screen and approaching the game screen attract more attention than other static objects, including stationary vehicles, red houses, and red billboards.


\subsection{Secondary Factors}\label{subsec::secondary factor}
In addition to the three key factors mentioned in \Cref{subsubsec::player's in-game speed,subsubsec::size and distance,subsubsec::motion states}, two secondary factors have relatively weak impacts on interest: the object's shapes and color contrast.


\subsubsection{Impacts of Uncommon Shape}\label{subsubsec::Uncommon Shapes}

There are two uncommon objects that attract players' focus. The first type of objects that can easily attract players' interest are those with uncommon physical features, such as unusual shapes. \cite{colorshape1_textureshape}. Another type is to point out objects in scenarios where they usually do not belong. 
Several examples of these types of unusual shapes are shown in \Cref{fig:shape}. The image in the top left shows a yacht on the road, while the one in the top right depicts a lake next to dry mountains. Both examples show objects appearing in unusual situations. The power tower in the bottom left image and the intersecting bridges in the right bottom image both have strange shapes.

\subsubsection{Impacts of High Contrast}\label{subsubsec::High Contrast}
\begin{figure}[t]
    \centering
    \includegraphics[width=0.8\linewidth]{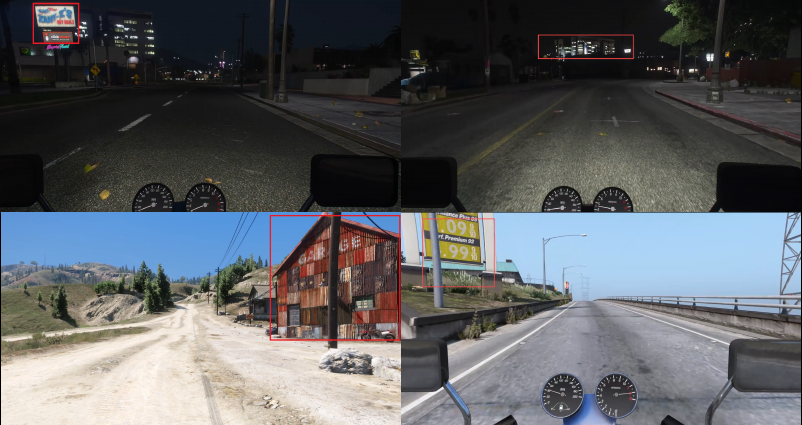}
    \caption{The top row of images shows high luminance contrast. The bottom row shows high hue contrast.}
    \label{fig:high contrast}
\end{figure}

High contrast means there is a significant difference in luminance or hue. \cite{Color_form_and_luminance_capture_attention,luminance_change_attention2012luminance}. Luminance contrast refers to the difference in brightness between an object and its surrounding environment. A typical scene involves bright billboards or buildings appearing in the night view, causing visual inconsistency. Hue contrast is the distinction between colors. Colors like red and green have a high hue contrast, whereas colors like blue and purple have a lower hue contrast.
Fig.~\ref{fig:high contrast} presents two types of instances with high contrast.
When an object has both color features and shape features simultaneously, it will have more ability to attract the player's interest. The enhanced ability can be simply added together, as color and shape are independent factors \cite{colorshape1_textureshape,colorshape2}. 

\section{Conclusion}\label{sec::conclusion}

We have created a cross-perspective first-person gaming dataset, containing images and videos. Each image has been annotated with multi-interest labels, and all objects within the images have been semantically segmented and processed to an easily accessible JSON file. We have analyzed the main factors and secondary factors that have an impact on the player's interest. The different player's in-game speed leads to significant differences in the distribution of object classes
In future work, we will quantitatively evaluate the impact of these factors on interest through rule-based methods or deep learning methods.

\section{Acknowledgments}
We would like to thank all the participants \textit{Yao Li, Chi Zhang, Lu Xu, Yuhao Cao, Kaiyao Shi, Bowen Tan, Changhua Wu, Yufan Yang, Mingxu Liu, Tao Li, Xinkun Liu, Tiecheng Jiang, Ying Zhang} in this experiment. It is their active participation, patience, and dedication that enable us to collect valuable data.

\bibliographystyle{IEEEtran}
\bibliography{IEEEabrv,ref}

\end{document}